\documentclass{pnastwo}
\usepackage{xcolor}
\usepackage{epsfig,amsmath,amssymb,graphicx,color,calc,epstopdf}
\usepackage{dsfont}
\usepackage{color}
\usepackage{amsmath}
\url{www.pnas.org/cgi/doi/}
\copyrightyear{2016}
\issuedate{Issue Date}
\volume{Volume}
\issuenumber{Issue Number}

\newcommand{\beq}{\begin{equation}} \newcommand{\eeq}{\end{equation}}
\usepackage{mathrsfs} \usepackage{graphicx,color} \usepackage{bbm} \usepackage{multirow}
\usepackage[normalem]{ulem}

\begin{document}

\title{Local structure can identify and quantify influential global spreaders in large scale social networks}

\author{Yanqing Hu\affil{1}{School of Data and Computer Science, Sun Yat-sen University, Guangzhou 510006, China}
Shenggong Ji\affil{2}{School of Information Science and Technology, Southwest  Jiaotong University, Chengdu 610031, China},
Yuliang Jin\affil{3}{CAS Key Laboratory for Theoretical Physics, Institute of Theoretical Physics, Chinese Academy of Sciences, Beijing 100190, China},
Ling Feng\affil{4}{Computing Science, Institute of High Performance Computing, A*STAR, 138632, Singapore}\affil{5}{Department of Physics, National University of Singapore, Singapore 117551},
H. Eugene Stanley\affil{7}{Center for Polymer Studies and Department of Physics, Boston University, Boston, MA 02215 USA},
Shlomo Havlin\affil{8}{Minerva Center and Department of Physics, Bar-Ilan University, Ramat Gan, Israell}
\contributor{Accepted by Proceedings of the National Academy of Sciences of the United States of America}
}


\significancetext{Identification and quantification of influential spreaders in social networks are challenging due to the gigantic network sizes and limited availability of the entire structure. Here we show that such difficulty can be overcome by reducing the problem scale to a local one, which is essentially independent of the entire network. This is because in viral spreading the characteristic spreading size do not depend on network structure outside the local environment of the seed spreaders. Our approach may open the door to solve various big data problems such as false information surveillance and control, viral marketing, epidemic control and network protection.
}

\maketitle

\begin{article}

\begin{abstract}

{Measuring and optimizing the influence of nodes in big-data online social networks are important for many practical applications, such as the viral marketing and the adoption of new products. As the viral spreading on social network is a global process, it is commonly believed that measuring the influence of nodes inevitably requires the knowledge of the entire network. Employing percolation theory, we show that the spreading process displays a nucleation behavior: once a piece of information spread from the seeds to more than a small characteristic number of nodes, it reaches a point of no return and will quickly reach the percolation cluster, regardless of the entire network structure; otherwise the spreading will be contained locally. Thus, we find that, without the knowledge of entire network, any nodes' global influence can be accurately measured using this characteristic number, which is independent of the network size. This motivates an efficient algorithm with constant time complexity on the long standing problem of best seed spreaders selection, with performance remarkably close to the true optimum.}

\end{abstract}



\dropcap{M}odern online social platforms are replacing traditional media~\cite{rust1994death} for the spreading of information and communication of opinions~\cite{kitsak2010identification,wangPeng2009Science,MM2015Nature, aral2012,bond201261}. A common feature of today's online social networks (OSNs) is their gigantic sizes -- for example, as of the second quarter of 2016, there are about $ 1.5$ billion monthly active users on Facebook. Noticeably, multiplicative explosions of some information may take place at a global scale in such gigantic OSNs, which is the foundation of  viral marketing strategies~\cite{ferguson2008word}. Because of this, quantification of viral spreading is traditionally believed to need global network information. Indeed, most measures,   such as  k-shell~\cite{kitsak2010identification}, degree discount~\cite{Chen2009}, cost-effective lazy forward~\cite{leskovec2007cost}, betweenness~\cite{Newman2001Betweenness}, closeness~\cite{freeman1979centrality}, Katz index~\cite{katz1953new}, evaluate the influence of nodes based on the knowledge of global network structures. In general, these methods become impractical for giant OSNs, because either the full network structural data is unavailable, or the computational time is non-scalable. On the other hand, based on massive social experiments, Christakis and Fowler proposed the so-called {\it three degrees of influence} (TDI) theory~\cite{CF2007,Fowler2013review}, which states that any individual's social influence ceases beyond three degrees (friends' friends' friends), and therefore suggests the existence of an unknown yet local effect. A recent study also shows that a local approximation works fairly well for a qualitative global measure of collective influence~\cite{MM2015Nature}. The above situation bares an apparent paradox, which inspires us to ask a fundamental question: could local network structure accurately determine the size of global spreading?  

\section{Result}

{\color{black}Here we recover a local characteristic infection size $s^*$ of the spreading process. It determines the key influence size in the stochastic spreading process described by the Susceptible-Infected-Recovered (SIR) family models~\cite{barrat2008dynamicalBook,colizza2006role,kempe2003maximizing,newman2002,castellano2009statistical,melonitrafficPNAS,eubank2004modelling}, which well describe the information spreading process in social media \cite{RevModPhysEpidemic,Nature1964aaEpid,ViralSpreadingPRE2011,RumerSpreading2008}. We find a ubiquitous {\color{black} and well separated,} bimodal behavior in the supercritical spreading regime - the spreading either extends {\it globally} reaching a finite fraction of the total population irrespective of the initial condition, or diminishes quickly beyond the {\it local} characteristic infection size (Fig~1A and 1C). The global and local phases are unambiguously separated. Using the mapping between the SIR family model and bond percolation \cite{newman2002,MendesRMP2008}, we provide a concrete physical understanding of these two well separated  phases. We show that the local phase {\color{black}can be used to accurately quantify} the node(s) spreading power (see Fig. 2A). In particular, the statistical properties of {\color{black}infected} cluster size distribution allow us to use solely local network structural information for selecting the best seed spreaders in significantly short constant time complexity.}

\section{Method}

Our study is carried out for SIR spreading mechanism on connected networks.
The central quantity of interest in the spreading model is the final number of activated nodes, or the spreading influence \cite{kempe2003maximizing}.
A common definition of the spreading influence of node $i$ is the expected number of active nodes originated from $i$:
\begin{equation}
S(i) \equiv \sum_{s=1}^N s \, g(i,s),
\label{eq:spreadability}
\end{equation}
where $g(i,s)$ is the probability that a total of $s$ nodes are eventually activated by node $i$ in a network of $N$ nodes, and the probability of an active node to activate a neighbouring node is $\beta$. In information spreading, an activated node corresponds to a spreader.
We find that the probability distribution function $g(i,s)$ has two prominent features: (i) It consists of two peaks, which correspond to local and viral  spreadings.  The local peak is located at small $s$, while the viral peak is centred at significantly larger $s$ (Fig.1A and C). Furthermore, the viral peak is a $\delta$ like-function, whose location is independent of node $i$ and different stochastic realizations~(SI. Sec. II).
 (ii) The two peaks are separated by a wide gap,
 which implies that one may introduce a small filtering size ,$ s^*$,  to distinguish between the two phases.

\begin{figure*}
\begin{center}
\includegraphics[width=0.9\linewidth]{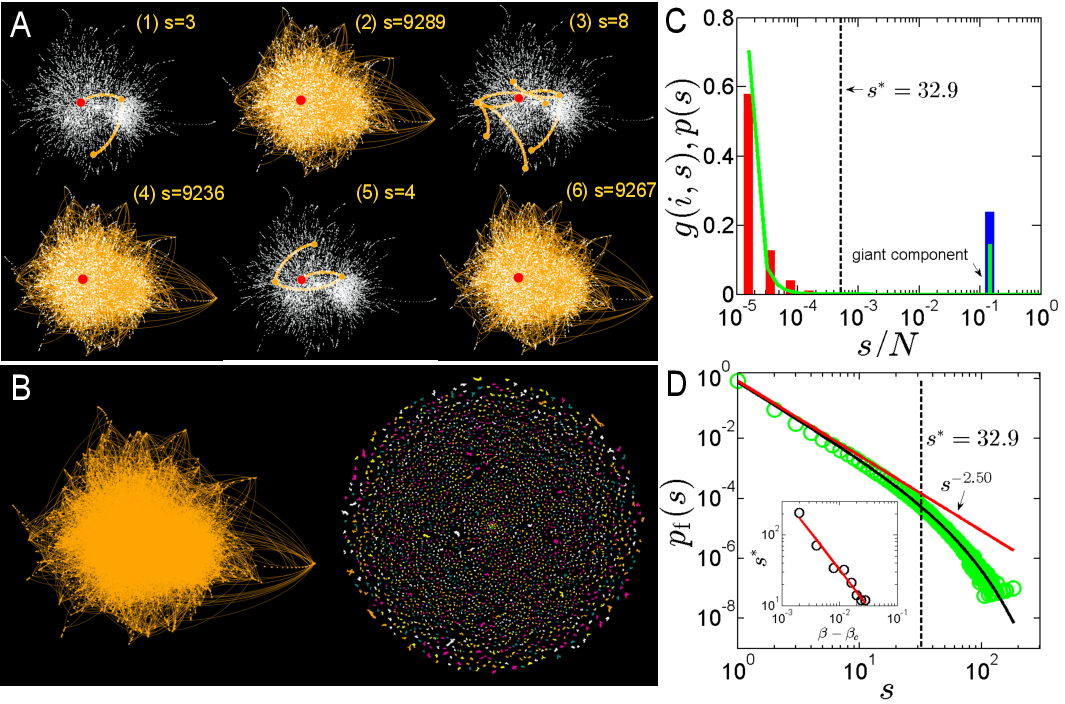}
\end{center}
\caption{{\bf Two phases phenomena.} (A) Examples of simulated local (1,3,5) and viral (2,4,6) SIR spreadings in the NOLA Facebook network ($\beta = 0.02, \beta_c = 0.01$). We start the simulation from a randomly chosen node (red, $k=27)$. The active and non-active nodes are colored in orange and white respectively.
(B) An illustration of giant (left) and finite (right) clusters in a bond percolation process.
 (C) The spreading probability distribution $g(i,s)$ (columns) is plotted together with the cluster size distribution function $p(s)$ (line) obtained from percolation. Note that $p(s)$ is the average of $g(i,s)$ over all nodes. In this example, we use the same seed node $i$ as in (A), but other randomly chosen nodes give similar bimodal distributions, with the same viral peak at $s^\infty$ (see SI Sec.~II). (D) The finite part $p_{\rm f}(s)$ (points) of $p(s)$  is fitted to Eq.~(\ref{eq:pf}) (black solid line) to obtain the characteristic size $s^* = 32.9 \pm 0.6$ and the exponent  $\tau = 2.50$.  (inset) The characteristic size $s^*$ is fitted to a power-law divergence near the critical point $\beta_c$, with a non-mean-field exponent $\sigma = 1.05$. The same network and the same $\beta$ are used in (A-D).}
\label{SpreadingAbility}
\end{figure*}

The statistical properties underlying these two features can be explored and better understood using the framework of percolation theory. This can be done through mapping the SIR process to bond percolation ~\cite{newman2002,castellano2009statistical},
where every link (bond) has a probability $1-\beta$ to be removed from the network (see SI Sec.~XII for the more general case where $\beta$ is link-dependent).
The final network forms many connected clusters of different sizes.  It has been proven that, the probability distribution function $g(i,s)$ in Eq.~(\ref{eq:spreadability}) is exactly equivalent to the cluster size distribution function $p(i,s)$, where $s$ is the size of the cluster that node $i$ belongs to~\cite{newman2002,newman2001random,HuPNAS1}. According to percolation theory, a giant component of size $s^\infty$ emerges above the percolation transition threshold $\beta_c$ (see Fig. 1B), where $p(i,s)$ is split into a finite (non-giant) part $p_{\rm f}(i, s)$ and a giant part $p(i,s^\infty)$ (Fig. 1C).
The size of $s^\infty$ is proportional to $N$, and depends on $\beta$.
Because $\sum s \,p_{\rm f}(i, s) \ll s^\infty p(i,s^\infty)$, we may approximate Eq.~(\ref{eq:spreadability}) as,
{\color{black}
\beq
S(i)\approx \hat{S}(i) \equiv  s^\infty \, p(i,s^\infty),
\label{eq:spreadability1}
\eeq
}
where $s^\infty = \sum_{i=1}^{N} p(i,s^\infty)$. In other words, {\it the spreading power {\color{black}of one node}
is the product of the giant component size and the probability that this node is in the giant component.} {\color{black} In information spreading, {\color{black}a broader type} of definition for `influence' exists by including  the {\color{black} nodes linked to the spreaders but do not {\color{black}spread the information} further, called `listeners' \cite{borge2012locating}. Since the underlying two-phase behavior {\color{black}is essentially} the same, {\color{black}the total number of listeners and spreaders increases monotonically with the total number of spreaders approximated in our percolation based algorithm. This implies that maximising the influence including listeners is equivalent to the problem of maximising the number of spreaders (SI. SEC.~III)}}}.

In artificial random networks with structure purely determined by the degree distributions, we can give the analytical solution for  this influence quantity  $\hat{S}(i)$ in Eq.~(\ref{eq:spreadability1}), with
\begin{equation}
p(i, s^\infty) = 1-(1-q)^{k_i},
\label{eq:GCProb}
\end{equation}
where $k_i$ is the degree of node $i$,  and $s^\infty = N \sum_{k=1}^{\infty} P(k) [1-(1-q)^k]$. Here, $q$ is the probability of a random link to be connected to the giant component and is determined from the self-consistent equation $q = \beta \sum_{k=1}^{\infty} \frac{kP(k)}{\langle k \rangle} [1-(1-q)^{k-1}]$, with average degree $\langle k \rangle$ and arbitrary degree distribution $P(k)$ \cite{Feng2015aa}.  The theoretical considerations and details for undirected, directed and degree-degree correlated are presented in the SI~Sec.~IV and~V.

For real networks whose structures are much more complex than random networks, an exact solution to the spreading influence is not possible. But the critical phenomenon and the statistical properties of the two phases remain the same (Fig. 1C). We can leverage on these properties, in particular the wide gap between these two phases to distinguish between viral and local spreadings, and construct methods to estimate the spreading influence of nodes in the network. In SIR processes, once the number of activated nodes reaches a threshold parameter $m$, the simulation could be terminated since this process is known to become most likely viral. We thus obtain a second approximated form for the node spreading power -- the truncated spreading power,
{\color{black}
\beq
S(i)\approx \tilde{S}(i) \equiv \tilde{s}^\infty \, \tilde{p}(i,\tilde{s}^\infty),
\label{eq:spreadability2}
\eeq
}
where $\tilde{p}(i,\tilde{s}^\infty)  \equiv \sum_{s=m}^{N} p(i,s)$, and $\tilde{s}^\infty  \equiv\sum_{i=1}^{N}  \tilde{p}(i,\tilde{s}^\infty)$. It turns out that percolation theory provides a fundamental insight into determining the threshold value $m$.  According to the theory, the distribution $ p_{\rm f}(i,s)$ has a fast decay tail $e^{-s/s^*}$,
where  $s^*$ gives a characteristic size of the finite components~\cite{newman2001random,BH1991book}.
For any $m \geq s^*$, the error introduced in $\tilde{S}(i)$ by truncating this tail  is  small (see Fig.~2A for a comparison between the real $S(i)$ and $\tilde{S}(i)$ in real networks). Figure 2B shows that the relative error $E^{\rm r}(i, m) \equiv [\tilde{S}(i) - S(i)]/S(i)$ decays quickly with $m$, and becomes negligible for $m \geq s^*$ (see SI~Sec.~IV and~V for a theoretical calculation of the error in random networks). {\color{black} Similar to the giant component size, the characteristic component size $s^*$ {\color{black}is} intrinsic to the whole network, and independent of the seed node $i$.} Hence the characteristic size $s^*$ has an important implication: once it is determined either theoretically or numerically, it can be used as a threshold for the parameter $m$. As long as $m$ is chosen to be above $s^*$, the truncated spreading power $\tilde{S}(i)$  is an excellent approximation for  $S(i)$, and its error is well controlled  (see SI. Sec.~VI).

\begin{figure}
\begin{center}
\includegraphics[width=0.9\linewidth]{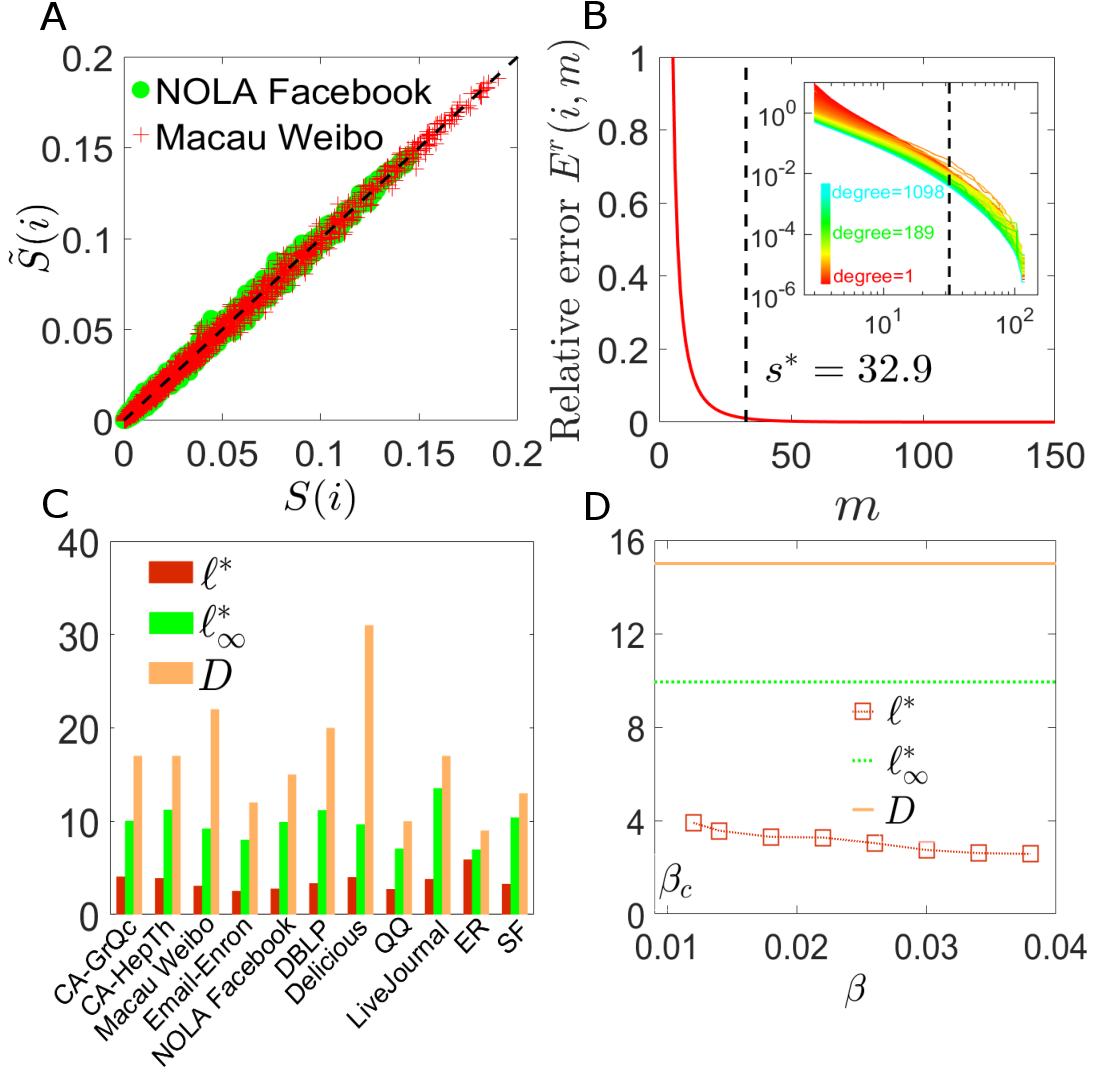}
\end{center}
\caption{{\bf Spreading power.}
(A) Comparison between the truncated spreading power $\tilde{S}(i)$ (Eq.~\ref{eq:spreadability2}) and the real exact spreading power $S(i)$ (Eq.~\ref{eq:spreadability}) in NOLA Facebook and Macau Weibo ($\beta_c = 0.05$) networks, where each point represents one node. (B) The $m$-dependence of the relative error $E^{\rm r}(i,m)$ of nodes whose degrees are equal to the average degree  $\langle k \rangle$, in the NOLA Facebook ($\beta = 0.02$) network.  (inset) The $m$-dependence of the relative error $E^{\rm r}(i,m)$ of nodes with different degrees.  The relative error decreases quickly with $m$ and becomes smaller than $1\%$ when $m > s^*$.  (C) {\color{black} Comparison among the influence radius $\ell^*$, the average distance of the furthest nodes from the seed nodes $\ell^*_{\infty}$ and the network diameter $D$ in nine OSNs, and two random networks (an ER network with $N=50000, \langle k \rangle =10$, and a scale-free (SF) network with $N=50000, P(k) \sim k^{-2.5}$).} We choose $\beta$ in different networks such that the fraction of the giant component is the same,  i.e., $s^\infty = 0.3 N$ (see SI. Sec.~I for the real networks description).  (D) The NOLA Facebook influence radius $\ell^*$ is smaller than both $\ell^*_{\infty}$ and $D$ for any $\beta > \beta_c$.
 }\label{error}
\end{figure}

The average of the cluster size distribution, $p(i,s)$, from seed node $i$ gives the global cluster distribution function $p(s) = \frac{1}{N} \sum_{i=1}^{N} p(i,s)$.
Excluding the giant component, its finite part $p_{\rm f}(s)$ has the same tail as that of  $p_{\rm f}(i,s)$~\cite{newman2001random,Shlomo2010Book,MendesRMP2008} (see Fig. 1D),
\beq
p_{\rm f}(s) \sim s^{-\tau} e^{-s/s^*},
\label{eq:pf}
\eeq
which can be used to obtain $s^*$ theoretically in random networks~\cite{newman2001random}.
For example, in an Erdos-Renyi (ER) network,  we obtain $s^{*}_{\rm ER}=\frac{1}{\beta \langle  k \rangle-1-\ln\beta-\ln \langle  k \rangle}$ (see SI Sec.~IV). An expansion of this expression around the percolation transition $\beta_c$ gives the critical scaling $s^{*} \sim |\beta - \beta_c|^{-1/\sigma} $, with the mean-field exponent $\sigma = 0.5$. For real OSNs, $s^*$ is obtained by fitting the simulation data to the exponential tail in Eq.~(\ref{eq:pf}) (see Fig. 1D and SI. Sec.~VII). Figure 1D inset shows that $s^*$ in real Facebook OSN also satisfies the critical power-law scaling.

To  reveal the topological meaning of the characteristic size $s^*$, we define an influence hopping radius $\ell^*$ associated to $s^*$. We perform SIR simulations until $s^*$ nodes are activated
and assign the maximum hopping distance (shortest path) between the seed and active nodes, averaged over all realizations and nodes, to be the influence radius $\ell^*$. For a typical $\beta$ such that $s^\infty = 0.3 N$,
we find that $\ell^* \sim 3-4$ in all OSNs studied,
which is significantly smaller than the average distance and diameter of the network as seen in Fig. 2C. This result shows that if an SIR spreading is local, then it would vanish within three to four steps, otherwise, it will spread to about $s^\infty = 0.3 N$ nodes. Note that $\ell$ increases when $\beta \to \beta_c$ (see Fig. 2D), whose scaling is discussed in SI Sec.~IV. This behaviour is analogous to a critical phenomena of a continuous phase transition:  at the critical point, the correlation length diverges, but as long as it moves beyond the critical point, a characteristic scale appears.

The above analysis resolves the seemingly paradox: while it is shown that the information spreading is in general a {\it global} process due to the viral spreading in the supercritical phase, the influence of any node basically only depends on its {\it local} network environment.{\color{black}While the computation time for $S(i)$  in Eq.~(\ref{eq:spreadability})  grows linearly with $N$, it
is reduced to a $N$-independent constant $O(m)$ for the truncated spreading power $\tilde{S}(i)$  in Eq.~(\ref{eq:spreadability2}). 
An important extension of this finding is that the method can be combined with many search algorithms for detecting the best spreaders and reduce their time complexity by  one order of $N$.}

\begin{figure*}
\begin{center}
\includegraphics[width=0.9\linewidth]{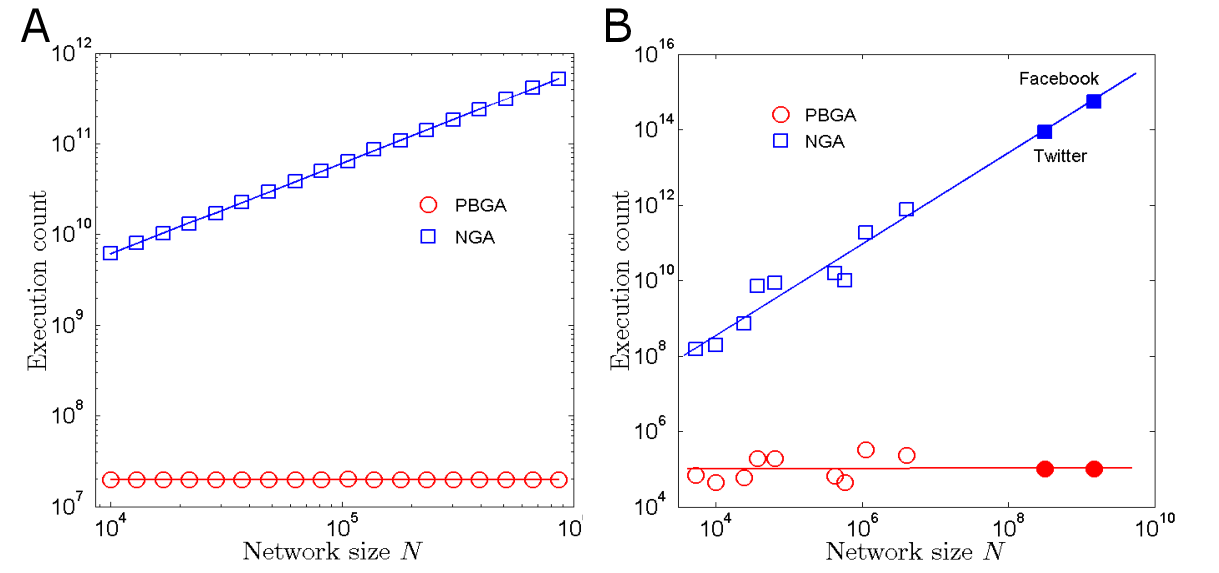}
\end{center}
\caption{{\bf Algorithm time complexity.} (A) Comparison of the computational execution count of percolation based greedy algorithm(PBGA) and natural greedy algorithm (NGA)\cite{Chen2009} in ER networks with $\beta = 0.2$ ($\beta_c = 0.1$). The algorithms select the set of $M=10$ most influential nodes out of $L=1000$ candidates with degree at $\langle k \rangle = 10$. Unlike natural greedy algorithm (NGA), PBGA's computational complexities is independent of network size.
(B) Comparison of the computational execution count (rescaled by $\langle k \rangle$ and $m$) of the same algorithms in real OSNs (open symbols, from left to right, CA-GrQc, CA-HepTh, Macau Weibo, Email-Enron, NOLA Facebook, DBLP, Delicous,
QQ, and LiveJournal, and please see SI. Sec.~I for the real networks description). {\color{black} The solid symbols are values extrapolated execution count based on the size of whole Twitter and Facebook networks.} The value of $\beta$ is chosen such that the giant component size is 30\% of the network size, i.e. $s^\infty =  0.3 N$ in each OSN.}
\label{MSpreaders}
\end{figure*}

Next, we aim to find the best $M$ spreaders $\mathcal{V}=\{v_1,v_2,\cdot \cdot \cdot, v_M\}$ from a given set $\mathcal{W}$  of $L$ candidates, to maximize their collective spreading power $S(\mathcal{V}) = \sum_{s=1}^N s \, p(\mathcal{V},s)$, where $p(\mathcal{V},s)$ is the probability that a total of $s$ nodes are activated by the selected spreaders in $\mathcal{V}$. Because it is usually more cost effective to target a large set of less influential nodes, rather than a small set of globally most influential nodes~\cite{bakshy2011}, we choose nodes with average properties (around average degree) as candidates. In practice, it is usually extremely difficult to obtain the full network structural information. Therefore unlike many other studies which select best seeds from the whole network, we only focus on a subset of candidate nodes. {\color{black}Extending from the formulation of a single node spreading power $\tilde{S}(i)$, we introduce a truncated collective spreading power $\tilde{S}(\mathcal{V}) \equiv \tilde{s}^\infty \, \tilde{p}(\mathcal{V},\tilde{s}^\infty)$, where  $\tilde{p}(\mathcal{V},\tilde{s}^\infty)$ is the probability that at least one cluster of at least $m$ nodes are activated by the $M$ seed spreaders. While the computation time for the collective influence $S(\mathcal{V})$  increases linearly with $N$ given any $\mathcal{V}$, it becomes $N$-independent for the estimator $\tilde{S}(\mathcal{V})$. }

Now we demonstrate one example of how to improve other algorithms and introduce new quantification capabilities through the combination of our approach with the natural greedy algorithm (NGA) \cite{kempe2003maximizing,Chen2009}. We call this algorithm {\it Percolation-based greedy algorithm (PBGA)} :
 (i) We first find the best spreader $\tilde{v}_1$ with the maximal individual spreading power based on the estimator $\tilde{S}(\tilde{v}_1)$,
(ii) then fix $\tilde{v}_1$ and find the second best spreader $\tilde{v}_2$ that maximizes the collective spreading power $\tilde{S} (\tilde{\mathcal{V}})$ for $\tilde{\mathcal{V}}=\{\tilde{v}_1, \tilde{v}_2\}$, and (iii) repeat this process $M$ times until $M$ spreaders $\tilde{\mathcal{V}} =\{\tilde{v}_1, \tilde{v}_2, \ldots \tilde{v}_M \}$ are selected. As a greedy algorithm, the PBGA maximizes the marginal gain in the objective function $\tilde{S}(\tilde{\mathcal{V}})$ at each step. Note that, replacing the objective function by the real spreading power in the above procedure would basically recover the NGA (see SI Sec.~IX for more details).

As expected, the simulation results show that the computational time in terms of execution count of PBGA is  {\color{black}independent} of network size $N$ (Fig. 3).
This reduction becomes significant for a world-wide online social network with billions of nodes. In SI~Table, we compare and summarize the theoretical time complexities of our percolation-based greedy algorithm, natural greedy algorithm and other widely used algorithms, including brute-force search, genetic algorithm, maximum degree, maximum k-shell \cite{kitsak2010identification}, {\color{black}degree discount heuristic~\cite{Chen2009},} maximum betweenness \cite{Newman2001Betweenness}, maximum closeness \cite{freeman1979centrality}, maximum Katz index \cite{katz1953new}, eigenvector method and  maximal collective influence (MCI) ~\cite{MM2015Nature}. {\color{black}Although, maximum degree, degree discount and MCI have $N$-independent theoretical computational complexities, the maximum degree and degree discount performance are much less than PBGA  and MCI is much slower than PBGA. This is because MCI needs the information of the nodes up to a distance $\ell$ of the seed nodes. In real networks which are small-world, a small $\ell$ would lead to thousands or more nodes. On the other hand, PBGA's complexity depends on $s^*$, which is independent of the small-world effect. }{\color{black}Fig.~S16 in the SI presents a graphical illustration of this difference.}

\begin{figure*}
\begin{center}
\includegraphics[width=0.8\linewidth]{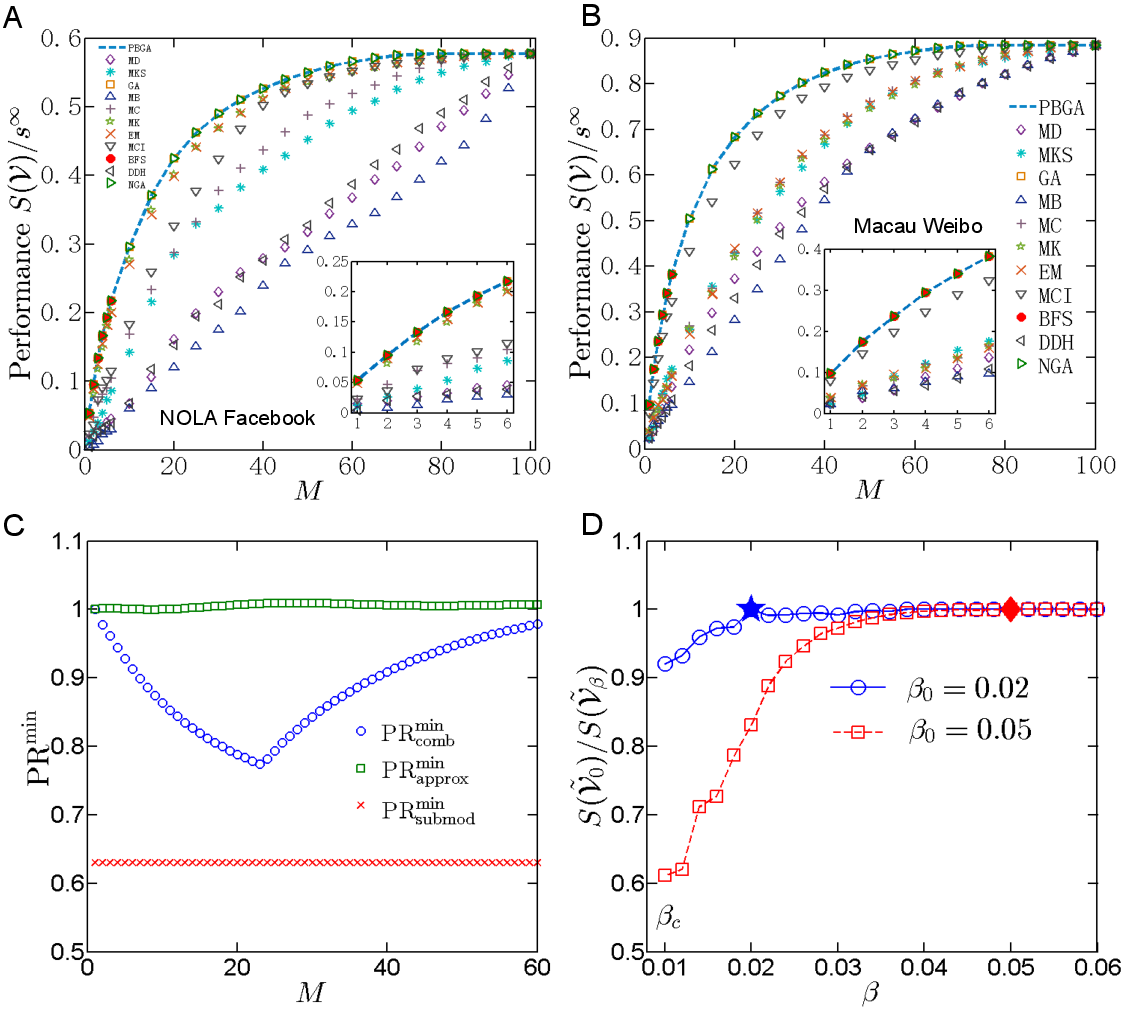}
\end{center}
\caption{{\bf Algorithm performance on real online social networks.} For (A)  NOLA Facebook ($\beta = 0.012$) and (B) Macau Weibo  ($\beta = 0.055$) network, we compare the algorithm performance of PBGA with other algorithms:  maximum degree (MD), natural greedy algorithm (NGA), brute-force search (BFS),  maximum k-shell (MKS)\cite{kitsak2010identification}, genetic algorithm (GA), maximum betweenness (MB) \cite{Newman2001Betweenness}, maximum closeness (MC) \cite{freeman1979centrality}, maximum Katz (MK) index \cite{katz1953new}, eigenvector method (EM), maximal collective influence (MCI)~\cite{MM2015Nature} and {\color{black}degree discount heuristic (DDH)~\cite{Chen2009}. The candidate nodes are randomly selected from the nodes with median degree nodes: degree is 10 for nodes in Facebook and out-degree is 3 for nodes from Weibo. Since the candidate nodes have the same degree, the maximum degree (MD) method is equivalent to the random selection of seed nodes.}
{\color{black} $S(\mathcal{V})$ is normalized by dividing the giant component size $s^\infty$.} Here with  $L=100$ candidates and $M$ varies from 1 to 100. (insets) The regime $1 \le M  \le 6$ is enlarged, where the rigorous  optimum obtained from BFS is available. (C) On the Facebook network, the combined rigorous lower bound ${\rm PR}^{\rm min}_{\rm comb}$, and the approximated lower bound ${\rm PR}^{\rm min}_{\rm approx}$, are plotted together with the sub-modular lower bound ${\rm PR}^{\rm min}_{\rm submod}=0.63$~\cite{kempe2003maximizing}, as functions of $M$. (D) {\color{black} The relative performance between PBGA solution based on $\beta=0.02$ ($\beta=0.05$) and the PBGA solution based on the other $\beta$ values, with both performances $\tilde{\mathcal{V}}_0$  and $\tilde{\mathcal{V}}_\beta$ estimated upon the same spreading rate $\beta$. The filled symbol's performance is exactly at $1$ since $\beta=\beta_0$. We see that as long as $\beta>\beta_0$, the solution at $\beta_0$ can be used at $\beta$ since their performances are almost the same, as long as both $\beta_0$ and $\beta$ are larger than the critical point $\beta_c\approx0.01$.
}}\label{MSpreaders}
\end{figure*}

We quantify the algorithm performance by comparing the collective spreading power $S(\mathcal{V})$ of the solution set $\mathcal{V}$ from different algorithms (Fig. 4A and 4B). Our results show that for the entire range of studied $M$,  the three algorithms,  PBGA, natural greedy algorithm and genetic algorithm, have the best  performances. Remarkably, the three algorithms give solutions indistinguishable from the true optimum obtained by brute-force algorithm, when $M$ is small (Fig. 4A and 4B inset). In particular, comparing the performance of PBGA and MCI in Fig. 5, we see that PBGA significantly outperforms MCI when the number $M$ of seed nodes is small. This can be understood since the original CI method \cite{MM2015Nature} deals with best nodes for breaking down the network, which are not necessarily the best spreaders. This is likely the reason behind the relatively lower performance of MCI (see more detailed discussion in SI. SEC. IX. B9). When $M$ becomes large, the performance difference diminishes, similar to the performance of any other algorithms as seen in Fig. 4A.
In fact, we conjecture that for any $M$, the solution of PBGA should be nearly optimal.

Another important aspect of this maximisation problem is to have a sense of how good is the solution compared to the true optimal solution $\mathcal{V}^*$, which is usually unknown \cite{kempe2003maximizing,leskovec2007cost}.
We give two lower bounds of the performance ratio ${\rm PR} \equiv S(\tilde{\mathcal{V}})/S({\mathcal{V}^*})$ between the PBGA performance $S(\tilde{\mathcal{V}})$ and the exact optimal performance $S(\mathcal{V}^*)$ (see Fig. 4C):
(i)
a combined bound ${\rm PR}^{\rm min}_{\rm comb} \equiv \max\{\frac{p(\mathcal{\tilde{V}},s^\infty)}{\sum_{i \in \mathcal{U}^*} p(i,s^\infty)}, \frac{p(\mathcal{\tilde{V}}, s^\infty)}{p(\mathcal{W}, s^\infty)}\}$, where $\mathcal{U}^*$ is the set of $M$ nodes with the maximum individual probability $p(i,s^\infty)$. It is rigorous for any networks. (ii) an approximated bound  ${\rm PR}_{\rm approx}^{\rm min} \equiv \frac{p(\mathcal{\tilde{V}}, s^\infty)}{1-\prod_{i \in \mathcal{U}^*}[1-p(i,s^\infty)]}$, that becomes rigorous in random networks (see SI. Sec.~X). The rigorous boundaries, $\frac{p(\mathcal{\tilde{V}},s^\infty)}{\sum_{i \in \mathcal{U}^*} p(i,s^\infty)}$ and $\frac{p(\mathcal{\tilde{V}}, s^\infty)}{p(\mathcal{W}, s^\infty)}$, work well in the small and large $M$ limits respectively, where they both approach to one, and the approximated bound ${\rm PR}_{\rm approx}^{\rm min} \approx 1$ for any $M$ value considered. Considering the above analysis,  we argue that PBGA gives a nearly optimized solution for an arbitrarily given number $M$ of spreaders.

In practical situations, the information spreading rate $\beta$ is usually unknown. However, our PBGA method could find close to optimal solutions  without knowing the exact $\beta$ value, as long as the information spreading is viral, i.e. supercritical region with $\beta>\beta_c$ (see SI. Sec. XI for the none-viral subcritical regions $\beta<\beta_c$). As illustrated in Fig.~4D, the solutions found at an arbitrary spreading rate $\beta_0$ performs nearly optimally at higher spreading rate $\beta>\beta_0$.  Thus without knowing the exact spreading rate, one can use a spreading rate slightly above the critical value $\beta_c$, such that the solutions perform optimally at  higher $\beta$ values.
On a related note, it has been observed that information spreading could exhibit bursty behaviours with different spreading speeds \cite{borge2012locating}. In such cases which could be mapped to having simultaneous different $\beta$ values over the course of spreading, possibly from small to large. This case still belongs to the SIR family. A more detailed analysis of this process is presented in SI Sec. XII.

\begin{figure}
\includegraphics[width=0.8\linewidth]{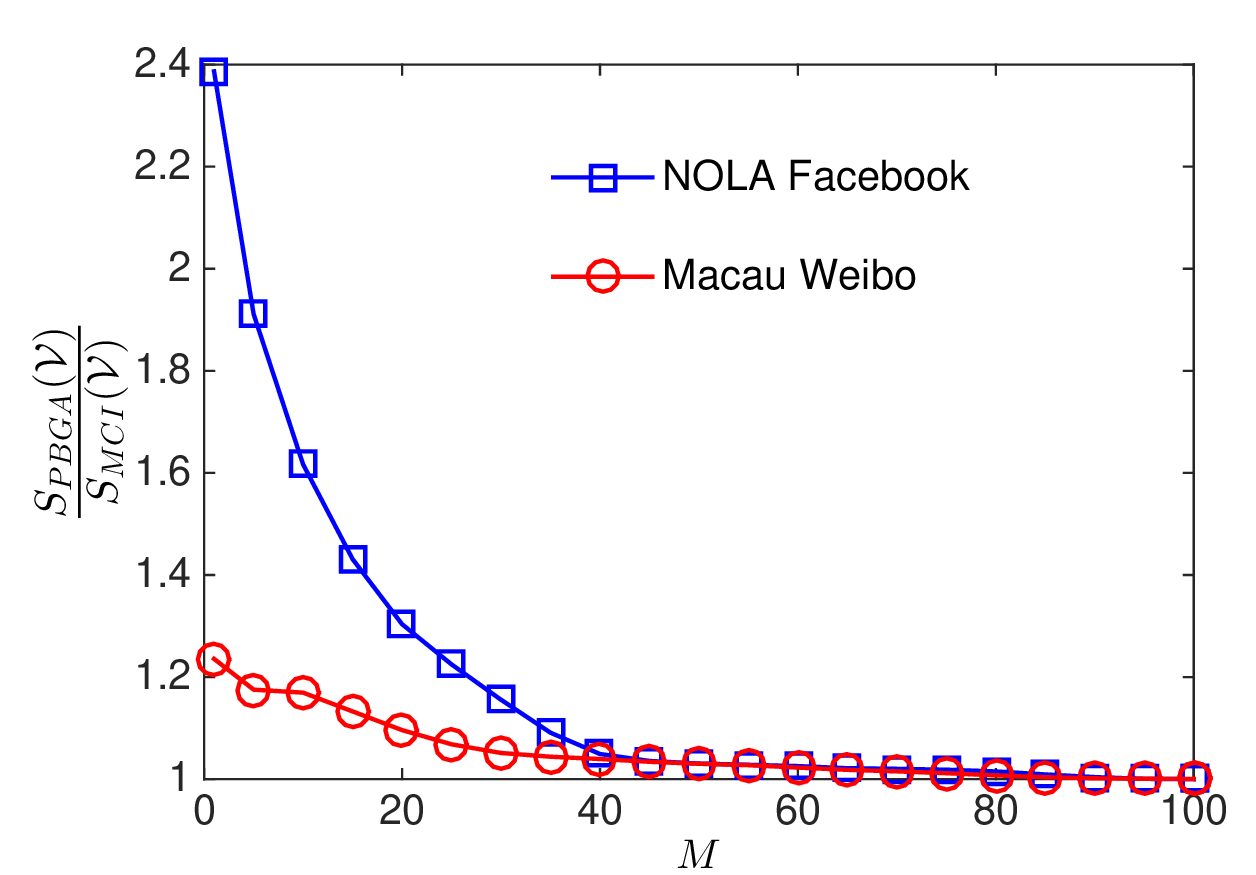}
\caption{{\bf Performance comparison between PBGA and MCI.} {\color{black} The vertical axis the the ratio between the seeds' influence of PBGA and MCI. For small number $M$ of seed nodes, PBGA significantly outperforms MCI. The difference diminishes as $M$ increases, when both solutions approach the theoretical maximum of giant component size. The simulation is carried out on the Facebook network with $\beta=0.012$.  } }
\label{ComparisonCI}
\end{figure}

\section{Summary}

In this work, we show from first principles that any node's influence can be quantified purely from its local network environment, based on the nature of the spreading dynamics. Our approach is distinct from other local attempts, which usually use some distance truncation strategies to approximate a relative global measure without the ability to quantify the actual influence. Although our framework is demonstrated on the basic SIR model, its applicability can be extended to several other spreading models if the following two properties hold: (i) For a collection of seed spreaders, the final steady state have two different outcomes of either being a localized outbreak with small and finite number of infections, or global epidemics with infectious/recovered population being proportional to the network size. (ii) When it is in the global outbreak, the size of the influence does not correlated with the initial spreader. See SI. SEC. XII  for discussions on a more general family of SIR models as well as for more complex models that include stiflers  \cite{borge2012absence} and SIS model \cite{RevModPhysEpidemic}.

\begin{acknowledgments}
We wish to thank the NSFC grant NO. 61773412, NO. 71731002, NO. U1711265, Guangzhou Science and Technology Project with NO. 201804010473 and Super computing application incubation project of SYSU.
\end{acknowledgments}

\bibliographystyle{unsrt}

\end{article}

\end{document}